\documentclass[doublecol]{epl2my}

\usepackage{hyperref}

\title{On the energy of particle collisions in the ergosphere of
the rotating black holes}
\shorttitle{On the energy of particle collisions in the ergosphere}

\author{A. A. Grib\footnote{E-mail: andrei\_grib@mail.ru}\inst{1,2,3} \and
Yu. V. Pavlov\footnote{E-mail: yuri.pavlov@mail.ru}\inst{1,2,4}}
\shortauthor{A. A. Grib, Yu. V. Pavlov}

\institute{
  \inst{1} A. Friedmann Laboratory for Theoretical Physics -
Saint Petersburg, Russia\\
  \inst{2} Copernicus Center for Interdisciplinary Studies -
Krak\'{o}w, Poland\\
  \inst{3} Theoretical Physics and Astronomy Department, The Herzen  University
- Moika 48, Saint Petersburg 191186, Russia\\
  \inst{4} Institute of Problems in Mechanical Engineering,
Russian Academy of Sciences -
Bol'shoy pr. 61, V.O., Saint Petersburg 199178, Russia
}

\abstract{
    It is shown that the energy in the centre of mass frame of colliding
particles in free fall at any point of the ergosphere of the rotating
black hole can grow without limit for fixed energy values on infinity.
    The effect takes place for large negative values of the angular momentum
of one of the particles.}

\begin{document}

\maketitle

%%%%%%%%%%%%%%%%%%%%%%%%%%%%%%%%%%%%%%%%%%%%%%%%%%%%%%%%%%%%%%%%%%%%%%
\section{Introduction}

    Recently many papers were
published~\cite{BanadosSilkWest09}--\cite{Zaslavskii12b} in which some
specific properties of collisions of particles close to the horizon of
the rotating black holes were discussed.
    In~\cite{BanadosSilkWest09} the effect of unlimited growing energy of
colliding two particles in the centre of mass frame for critical Kerr
rotating black holes (call it BSW effect) was discovered.
    In our papers~\cite{GribPavlov2010}--\cite{GribPavlov2011b}
the same effect was found for noncritical black holes when multiple
collisions took place.

    All this shows that close to the horizon there is natural supercollider
of particles with the energies up to Planck's scale.
    However its location close to the horizon makes difficult due to the large
red shift get some observed effects outside of the ergosphere when particles
go from the ``black hole supercollider'' outside.
    Here we shall discuss some other effect valid at any point of
the ergosphere.
    The energy in the centre of mass frame can take large values for large
negative values of the angular momentum projection on the rotation axis of
the black hole.
    The problem is how to get this large (in absolute value) negative values.
    That is why first we shall obtain limitations on the values of
the projection of angular momentum outside ergosphere for particles falling
into the black hole and inside the ergosphere.
    It occurs that inside the ergosphere of black hole there is no limit for
negative values of the momentum and arbitrary large energy in the centre
of mass frame is possible for large angular momentum of the two colliding
particles.
    In a sense the effect is similar to the BSW-effect.
    One of the particle with large negative angular momentum can be called
``critical'', the other particle ``ordinary''.

    The system of units $G=c=1$ is used.

%%%%%%%%%%%%%%%%%%%%%%%%%%%%%%%%%%%%%%%%%%%%%%%%%%%%%%%%%%%%%%%%%%%%%%
\section{General formulas for the collision energy close
to the black hole}
\label{sec2}

    The Kerr's metric of the rotating black hole~\cite{Kerr63} in
Boyer--Lindquist coordinates~\cite{BoyerLindquist67} has the form:
    \begin{eqnarray}
d s^2 &=& d t^2 -
\frac{2 M r}{\rho^2} \, ( d t - a \sin^2 \! \theta\, d \varphi )^2
\nonumber \\
&&-\, \rho^2 \left( \frac{d r^2}{\Delta} + d \theta^2 \right)
- (r^2 + a^2) \sin^2 \! \theta\, d \varphi^2,
\label{Kerr}
\end{eqnarray}
    where
    \begin{equation} \label{Delta}
\rho^2 = r^2 + a^2 \cos^2 \! \theta, \ \ \ \ \
\Delta = r^2 - 2 M r + a^2,
\end{equation}
    $M$ is the mass of the black hole, $ aM $ its angular momentum.
    The rotation axis direction corresponds to $\theta =0$, i.e. $a \ge 0$.
    The event horizon of the Kerr's black hole corresponds to
    \begin{equation}
r = r_H \equiv M + \sqrt{M^2 - a^2} .
\label{Hor}
\end{equation}
    The surface of the static limit is defined by
    \begin{equation}
r = r_1(\theta) \equiv M + \sqrt{M^2 - a^2 \cos^2 \theta} .
\label{Lst}
\end{equation}
    In case $ a \le M $ the region of space-time between the static limit
and event horizon is called ergosphere.

    For geodesics in Kerr's metric~(\ref{Kerr}) one obtains
(see \cite{Chandrasekhar}, Sec.~62 or \cite{NovikovFrolov}, Sec.~3.4.1)
    \begin{equation} \label{geodKerr1}
\rho^2 \frac{d t}{d \lambda } = -a \left( a E \sin^2 \! \theta - J \right)
+ \frac{r^2 + a^2}{\Delta}\, P,
\end{equation}
    \begin{equation}
\rho^2 \frac{d \varphi}{d \lambda } =
- \left( a E - \frac{J}{\sin^2 \! \theta} \right) + \frac{a P}{\Delta} ,
\label{geodKerr2}
\end{equation}
    \begin{equation} \label{geodKerr3}
\rho^2 \frac{d r}{d \lambda} = \sigma_r \sqrt{R}, \ \ \ \
\rho^2 \frac{d \theta}{d \lambda} = \sigma_\theta \sqrt{\Theta},
\end{equation}
    \begin{equation} \label{geodP}
P = \left( r^2 + a^2 \right) E - a J,
\end{equation}
    \begin{equation} \label{geodR}
R = P^2 - \Delta [ m^2 r^2 + (J- a E)^2 + Q],
\end{equation}
    \begin{equation} \label{geodTh}
\Theta = Q - \cos^2 \! \theta \left[ a^2 ( m^2 - E^2) +
\frac{J^2}{\sin^2 \! \theta} \right].
\end{equation}
    Here $E$ is conserved energy (relative to infinity)
of the probe particle,
$J$ is conserved angular momentum projection on the rotation axis
of the black hole,
$m$ is the rest mass of the probe particle, for particles with nonzero
rest mass $\lambda = \tau /m $,
where $\tau$ is the proper time for massive particle,
$Q$ is the Carter's constant.
    The constants $\sigma_{r}, \sigma_{\theta}$ in
formulas~(\ref{geodKerr3}) are equal $\pm 1$ and are defined by the direction
of particle movement in coordinates $r$, $\theta$.
    For massless particles one must take $m = 0$
in~(\ref{geodR}), (\ref{geodTh}).

    One can find the energy in the centre of mass frame of two colliding
particles $E_{\rm c.m.}$ with rest masses~$m_1$, $m_2$ taking the square of
    \begin{equation} \label{SCM}
\left( E_{\rm c.m.}, 0\,,0\,,0\, \right) = p^{\,i}_{(1)} + p^{\,i}_{(2)},
\end{equation}
    where $p^{\,i}_{(n)}$ are 4-momenta of particles $(n=1,2)$.
    Due to $p^{\,i}_{(n)} p_{(n)i}= m_n^2$ one has
    \begin{equation} \label{SCM2af}
E_{\rm c.m.}^{\,2} = m_1^2 + m_2^2 + 2 p^{\,i}_{(1)} p_{(2)i} .
\end{equation}
    Note that the energy of collisions of particles in the centre of mass
frame is always positive (while the energy of one particle due to Penrose
effect~\cite{Penrose69} can be negative!) and satisfies the condition
    \begin{equation} \label{Eb0}
E_{\rm c.m.} \ge m_1 + m_2.
\end{equation}
    This follows from the fact that the colliding particles move one
towards another with some velocities.

    It is important to note that $E_{\rm c.m.}$ for two colliding particles
is not a conserved value differently from energies of particles (relative
to infinity) $E_1$,  $E_2$.

    For the free falling particles with energies $E_1$,  $E_2$ and angular
momentum projections $J_1, J_2$ from~(\ref{geodKerr1})--(\ref{geodR})
one obtains~\cite{HaradaKimura11}:
    \begin{eqnarray}
E_{\rm c.m.}^{\,2} &=& m_1^2 + m_2^2 +
\frac{2}{\rho^2} \biggl[ \, \frac{P_1 P_2 -
\sigma_{1 r} \sqrt{R_1} \, \sigma_{2 r} \sqrt{R_2}}{\Delta}
\nonumber \\
&&- \, \frac{ (J_1 - a  E_1 \sin^2 \! \theta) (J_2 - a  E_2 \sin^2 \! \theta)}
{\sin^2 \! \theta}
\nonumber \\
&& -\, \sigma_{1 \theta} \sqrt{\Theta_1} \,
\sigma_{2 \theta} \sqrt{\Theta_2} \, \biggr].
\label{KerrL1L2}
\end{eqnarray}

%%%% *****************************************************************
\section{Limitations on the values of particle
angular momentum close to the Kerr's black hole}
\label{sec3}

    The permitted region for particle movement is defined by conditions
    \begin{equation} \label{ThB0}
\Theta = Q - \cos^2 \! \theta \left[ a^2 ( m^2 - E^2) +
\frac{J^2}{\sin^2 \! \theta} \right] \ge 0,
\end{equation}
    $R \ge 0$,
and movement ``forward in time'' leads to
$ d t / d \lambda >0$~\cite{Wald}.
    The condition~(\ref{ThB0}) gives possible values of Carter's constant~$Q$.
    From~(\ref{geodKerr3}) it follows that for movement with constant $\theta$
it is necessary and sufficient that $\Theta=0$.
    It is always possible to choose this value,
so that for geodesics in the equatorial plane $(\theta=\pi/2)$ Carter's
constants $Q=0$.

    Let us find limitations for the particle angular momentum from the
conditions $R \ge 0$, $ d t / d \lambda >0$ at the point $(r, \theta)$,
taking the fixed values of $\Theta$.
    Outside the ergosphere $ r^2 -2 r M +a^2 \cos^2 \! \theta >0 $ one obtains
    \begin{equation} \label{EvErg}
E \ge \frac{1}{\rho^2} \sqrt{(m^2 \rho^2 + \Theta)
(r^2 -2 r M +a^2 \cos^2 \! \theta)},
\end{equation}
    \begin{equation} \label{JvErg}
J \in \left[ J_{-}, \ J_{+} \right],
\end{equation}
    \begin{eqnarray}
J_{\pm} = \frac{\sin \theta}{r^2 -2 r M +a^2 \cos^2 \! \theta}
\biggl[ - 2 r M a E \sin \theta
\nonumber \\
\pm \, \sqrt{ \Delta \left( \rho^4 E^2 - (m^2 \rho^2 + \Theta)
(r^2 -2 r M +a^2 \cos^2 \! \theta) \right)} \biggr] .
\label{Jpm}
\end{eqnarray}
    Boundary values of $ J_{\pm} $ correspond to $ d r/d \lambda = 0 $.

    On the boundary of ergosphere
    \begin{equation} \label{rEgErg}
r = r_1(\theta) \ \ \ \Rightarrow \ \ \ E \ge 0,
\end{equation}
    \begin{equation} \label{JgErg}
J \le E \left[ \frac{M r_1(\theta) }{a} + a \sin^2 \! \theta \left(
1 - \frac{m^2}{2 E^2} - \frac{\Theta}{4 M r_1(\theta) E^2} \right) \right].
\end{equation}

    Inside ergosphere
    \begin{equation} \label{lHmdd}
r_H < r < r_1(\theta) \ \ \ \Rightarrow \ \ \
(r^2 -2 r M +a^2 \cos^2 \! \theta) <0 ,
\end{equation}
    \begin{eqnarray}
J \le J_{-}(r,\theta) = \frac{\sin \theta}{
-(r^2 \!-\! 2 r M \!+\! a^2 \cos^2 \! \theta)} \biggl[ 2 r M a E \sin \theta
\nonumber \\
- \, \sqrt{ \Delta \left( \rho^4 E^2 - (m^2 \rho^2 + \Theta)
(r^2 -2 r M +a^2 \cos^2 \! \theta) \right)} \biggr] .
\label{JmErg}
\end{eqnarray}
    So on the {\it boundary and inside ergosphere there exist geodesics on
which particle with fixed energy can have arbitrary large in absolute value
negative angular momentum projection.}

    From~(\ref{JmErg}) one can see that for negative energy~$E$ of the particle
in ergosphere its angular momentum projection on the rotation axis of the
black hole must be also negative.
    This is a well known Penrose effect~\cite{Penrose69}.
    However rotation in Boyer--Lindquist coordinates for any particle in
ergosphere has the same direction as the rotation of the black hole
(the effect of the ``dragging'' of bodies by the rotating black
hole~\cite{NovikovFrolov}).
    Really for timelike geodesics $d s^2 >0$ leads to $d \varphi / d t >0$.
    So it  is incorrect to say (as it is said in some test books on black
holes~\cite{Chandrasekhar}, p.~368) that
``only counter-rotating particles can have negative energy''!
    Inside the ergosphere the usual intuition which is true far outside it
that the change of the sign of angular momentum projection on $Z$-axis means
the change of the rotation on counter-rotation following from usual
formula $\mathbf{J} = \mathbf{r} \times \mathbf{p} $ is incorrect.

    From equations of geodesics~(\ref{geodKerr2}) one can see the peculiar
properties of the correspondence between the direction of rotation and
the angular momentum projection.
    Let us rewrite it as
    \begin{equation} \label{KerrRotat}
\rho^2 \sin^2 \! \theta \frac{d \varphi}{d \lambda} =
\frac{E 2 M r a \sin^2 \! \theta}{\Delta} + J \,
\frac{r^2 - 2 r M + a^2 \cos^2 \! \theta}{\Delta}.
\end{equation}
    For large $r$ outside the ergosphere one gets the standard expression
for the angular momentum projection in Minkowski space
$J= m r^2 \sin^2 \! \theta\, d \varphi / d \tau$
and $J= m r^2 \, d \varphi / d \tau$ for $\theta = \pi/2$.

    But when one is ingoing inside the ergosphere %% from~(\ref{KerrRotat})
one sees that the coefficient for $J$ in~(\ref{KerrRotat})
becomes zero on its surface so that the angular velocity
$ d \varphi / d \lambda $ and $ d \varphi / d \tau $ is defined by the energy
and does not depend on $J$ at all.

    Inside the ergosphere the coefficient for $J$ in ~(\ref{KerrRotat})
becomes negative, the angular velocity is still positive and one comes
to unusual conclusion:
{\it if the energy of the particle in ergosphere is fixed particles with
negative angular momentum projection are rotating in the direction of
rotation of the black hole with greater angular velocity}!

    So the constant characterizing the geodesic which coincides with the
usual angular momentum definition taken from Newtonian physics outside
the black hole does not coincide with it in ergosphere.

%%%% *****************************************************************
\section{Energy of collision with a particle with large angular momentum}
\label{secErgo}

    Let us find the asymptotic of~(\ref{KerrL1L2}) for $J_2 \to -\infty$ and
some fixed value $r$ in ergosphere supposing the value of Carter's
constant $Q_2$ to be such that~(\ref{ThB0}) is valid and $\Theta_2 \ll J_2^2$.
    Then from~(\ref{KerrL1L2}) one obtains
    \begin{eqnarray}
\hspace*{-32pt}
E_{\rm c.m.}^{\,2} \approx \frac{- 2 J_2}{\rho^2 \Delta } \,
\biggl[ \frac{J_1}{\sin^2 \! \theta}
\left( r^2 -2 r M +a^2 \cos^2 \! \theta \right)
\nonumber \\ +\, 2 r M a E_1
- \frac{\sigma_{1r} \sigma_{2r} \sqrt{R_1}}{\sin \theta}
\sqrt{-(r^2 \!-\! 2 r M \!+\! a^2 \cos^2 \! \theta) } \biggr] .
\label{KerrJB}
\end{eqnarray}
    This asymptotic formula is valid for all possible $E_1$, $J_1$
(see~(\ref{JmErg})) for $r_H < r < r_{1}(\theta)$ and for
$E_1>0$ and $J_1$ satisfying~(\ref{JgErg}) for $r=r_{1}(\theta)$.
    The poles $\theta = 0, \pi$ are not considered here because the points
on surface of static limit are on the event horizon.

    Note that expression in brackets in~(\ref{KerrJB}) is positive
in ergosphere.
    This is evident for $r=r_{1}(\theta)$ and follows from
limitations~(\ref{JmErg}) for $r_H < r < r_{1}(\theta)$,
and inside ergosphere~(\ref{KerrJB}) can be written as
    \begin{eqnarray}
E_{\rm c.m.}^{\,2} \approx J_2 \frac{r^2 -2 r M +a^2 \cos^2 \! \theta}
{\rho^2 \Delta \sin^2 \! \theta}
\nonumber \\
\times \left( \sigma_{1r} \sqrt{J_{1 +}- J_1} -
\sigma_{2r} \sqrt{J_{1 -}- J_1} \right)^2.
\label{KerrJBner}
\end{eqnarray}

    So from~(\ref{KerrJB}) one comes to the conclusion that
{\it when particles fall on the rotating black hole collisions with arbitrarily
high energy in the centre of mass frame are possible at any point of
the ergosphere if $J_2 \to -\infty$ and the energies $E_1, E_2$ are fixed}.
    The energy of collision in the centre of mass frame is growing
proportionally to $\sqrt{|J_2|}$.

    Note that for large $-J_2$ the collision energy close to horizon
in the centre of mass frame depending on values $E_1, J_1$
can be as large as less then for collisions at the other points of ergosphere.

\begin{figure}[h]
\onefigure[width=77mm]{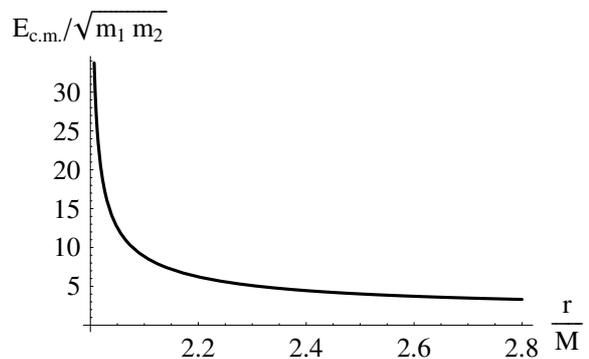}
\caption{The collision energy in the centre of mass frame for particles with
$J_1=0$ and $J_2=J_{-}$ out of the ergosphere.}
\label{FigEnJp}
\end{figure}

    Outside ergosphere the collision energy is limited for given $r$, but for
$ r \to r_1$ it can be large if one of the particles gets in intermediate
collisions the angular momentum close to $J_{-}$~(see~(\ref{Jpm})).
    On fig.~\ref{FigEnJp} the dependence of collision energy in the centre
of mass frame on the coordinate $r$ is shown for particles with
$ E_1 = E_2 = m_1 = m_2 $, $J_1=0$ and $J_2=J_{-}$
moving in equatorial plane of black hole with $ a=0.8 M $.
%%%%%%%%%%%%%%%%%%%%%%%%%%%%%%%%%%
%%%%%%%%%%%%%%%%%%%%%%%%%%%%%%%%%%

    Note that large negative values of the angular momentum projection are
forbidden for fixed values of energy of particle out of the ergosphere.
    So particle which is nonrelativistic on space infinity ($E=m$)
can arrive to the horizon of the black hole if its angular momentum projection
is located in the interval
    \begin{equation}
-2 m M \left[ 1 + \sqrt{1+ \frac{a}{M}}\, \right] \le J \le 2 m M
\left[ 1 + \sqrt{1 - \frac{a}{M}}\, \right].
\label{KerrEM}
\end{equation}
    The left boundary is a minimal value of the angular momentum of particles
with $E=m$ capable to achieve ergosphere falling from infinity.
    That is why collisions with $J_2 \to -\infty$ do not occur for particles
following from infinity.
    But if the particle came to ergosphere and there in the result of
interactions with other particles is getting large negative values of the
angular momentum projection (no need for getting high energies!)
then its subsequent collision with the particle falling on the black hole
leads to high energy in the centre of mass frame.

    Getting superhigh energies for collision of usual particles (i.e. protons)
in such mechanism occur however physically nonrealistic.
    Really from~(\ref{KerrJB}) the value of angular momentum necessary for
getting the collision energy $E_{\rm c.m.}$ has the order
    \begin{equation}
J_2 \approx - \frac{a E_{\rm c.m.}^{\,2}}{2 E_1}.
\label{KerrEMR}
\end{equation}
    So from~(\ref{KerrEM}) absolute value of the angular momentum $J_2$
must acquire the order $ E_{\rm c.m.}^{\,2} / (m_1 m_2)$ relative to the
maximal value of the angular momentum of the particle incoming to ergosphere
from infinity.
    For example if $E_1=E_2 = m_p$ (the proton mass) then $|J_2|$ must
increase with a factor $10^{18}$ for $ E_{\rm c.m.} = 10^9 m_p$.
    To get this one must have very large number of collisions with getting
additional negative angular momentum in each collision.

    However the situation is different for supermassive particles.
    In~\cite{GribPavlov2002(IJMPD)}--\cite{GribPavlov2008c} we discussed
the hypothesis that dark matter contains stable superheavy neutral particles
with mass of the Grand Unification scale created by gravitation in the end
of the inflation era.
    These particles are nonstable for energies of interaction of the order
of Grand Unification and decay on particles of visual matter but are
stable at low energies.
    But in ergosphere of the rotating black holes such particles due to
getting large relative velocities can increase their energy from $2m$ to
values of $3m$ and larger so that the mechanism considered in our paper
can lead to their decays as it was in the early universe.
    The member of intermediate collisions for them is not very large
(of the order of 10).

%%%%%%%%%%%%%%%%%%%%%%%%%%%%%%%%%%%%%%%%%%%%%%%%%%
\acknowledgments
The research is supported by the grant from The John Templeton Foundation.

%%%% ****************************************************************

\end{document}